

\magnification=\magstep1
\vsize = 23.5 truecm
\hsize = 15.5 truecm
\hoffset = .2truein
\baselineskip = 14 truept
\nopagenumbers


\def\rr { {\bf r } }

\topinsert
\vskip 4 truecm
\endinsert

\centerline{\bf TIME DEPENDENT DENSITY FUNCTIONAL THEORY}
\centerline{\bf IN QUANTUM LIQUIDS}

\vskip 16 truept

\centerline{\bf Sandro Stringari}

\vskip 5 truept

\centerline{Dipartimento di Fisica, Universit\`a di Trento }
\centerline{I-38050 Povo, Trento, Italy }

\vskip 2 truecm


\centerline{\bf 1.  INTRODUCTION}
\vskip 12 truept

The density functional (DF) approach is a well established theory for
investigating important ground state properties (energies and density
profiles) of quantum many body systems.
In the last decades it has been systematically applied
(mainly in the so called local density approximation) to a large variety of
electronic systems. Applications to strongly interacting quantum systems
(helium) has been also become popular in the recent years providing a useful
and stimulating alternative (see for example [1]) to microscopic
{\it ab initio} calculations of inhomogeneous systems (free surface, films,
clusters ...).

The equations of DF theory for static calculations are derived from a
variational principle of the form
$$
\delta \left( E - \epsilon_i \int \! d\rr \ \Psi_i^* \Psi_i \right) \ = \ 0
\eqno(1)
$$
where
$$
E = \int \! d\rr  \ {\cal H}
\eqno(2)
$$
is the energy functional. The energy density ${\cal H}$, characterizing the
functional, in general depends on the 1-body density matrix (diagonal as well
as non diagonal components). The  equations of DF theory have the form of
Hartree (or Hartree-Fock) equations for fermions and of an Euler equation
for bosons.

It is important to recall  the relevant  physical quantities
that DFT should account for. Concerning bulk properties they are:
\smallskip
\item{$\bullet$} The equation of state
\smallskip
\item{$\bullet$} The static response function.

The equation of state is fixed by the knowledge of ${\cal H}$ in the
homogeneous
limit and in particular by its density dependence according to the law
$$
P = \rho^2 {\partial \over \partial \rho} {{\cal H} \over \rho} \ \ \ .
\eqno(3)
$$

The static response function  is a key quantity of the theory that should be
(in principle) {\it exactly} accounted for  not only in the
macroscopic regime where it coincides with the compressibility,
but also  at higher wave vectors where the response of the system is
sensitive to
microscopic details of correlations among particles.

In DFT the static response $\chi(q)$ corresponds to the Fourier transform
of the second derivative of the energy functional with respect to the density
$$
\chi =   {\delta^2 E \over \delta \rho({\bf r}_1) \delta \rho({\bf r}_2)}
\ \ \ .
\eqno(4)
$$
The static  response is related to the dynamic structure factor through the
well known equation [1]
$$
\chi(q) = -2\int^{\infty}_0 \! d\omega \  \omega^{-1} S(q,\omega) \ \ \ .
\eqno(5)
$$
This equation reveals an important and useful connection between static
($\chi(q,\omega)$) and dynamic ($S(q,\omega)$) features of the system.

In superfluid $^4$He both the equation of state and the static response
are well known experimentally in a rather wide range of
densities and wave vectors respectively.
The most refined phenomenological density functional recently proposed by the
Orsay-Trento team accounts with good accuracy for the fine details of
these quantities [2] (see figs. 1 and 2).
\topinsert
\vskip 11 truecm
\noindent
{\bf Figure 1.}   Equation of state of superfluid $^4$He. The full line
corresponds
to the density functional of ref.[2]. The dashed line is the prediction
of the Monte Carlo calculations of ref.[3]. Points are the experimental
values (from ref.[4]).
\vskip 28 truept
\endinsert

\topinsert
\vskip 11 truecm
\noindent
{\bf Figure 2.} Static response function of superfluid $^4$He. The full line
 corresponds to the density functional of ref.[2]. Experimental points are
from ref. [5].
\vskip 28 truept
\endinsert
It is also important to recall that DFT is not suitable to account for
all the ground
state properties of the system. In particular the pair correlation function
and hence its
 Fourier transform, the static structure factor $S(q)$,  lies
outside the predictive power of the theory.
At zero temperature the quantity $S(q)$ is related to the dynamic structure
factor through the most famous relation
$$
S(q) =  \int^{\infty}_0 d\omega  S(q,\omega) \ \ \ .
\eqno(6)
$$
Attempts to force the theory  to account for the exact value of $S(q)$,
through a suitable parametrization of the dynamic structure factor
$S(q,\omega)$), should be considered unphysical.

The formalism of DFT can be easily extended to the time dependent case (TDDFT).
However, while in the static case the theory
is well established and is expected
to be, at least in principle, an exact approach
for the determination of energy and density
profiles of inhomogeneous systems, the
applicability of TDDFT to the investigation of dynamic phenomena (for
example propagation of collective phenomena) is less obvious especially in the
microscopic regime of large wave vectors.
The purpose of this paper is to discuss some relevant questions concerning
the {\bf structure}, the {\bf ingredients} and the {\bf applicability} of
TDDFT. Some emphasis will be also given to establish important
connections between
TDDFT and microscopic theories such as the Feynman theory for the elementary
excitations of Bose superfluids.

The equations of TDDFT are naturally derived starting from the least action
principle
$$
\delta \int_{t_1}^{t_2} dt \int \! d\rr \left[ {\cal H}
-\epsilon_i \Psi_i^* \Psi_i - \Psi_i^* i  {\partial  \over  \partial t }
\Psi_i  \right] \ =\ 0
\eqno (7)
$$
and take the form of a  Schr\"odinger-like equation
$$
(\tilde{H} - \epsilon_i )  \Psi_i = i  {\partial \over \partial t } \Psi_i
 \ \ \ ,
\eqno (8)
$$
where $\tilde{H}$ is a 1-body density dependent hamiltonian to be determined,
together with the solution of the Schr\"odinger equation, in a selfconsistent
way.

In the case of bosons, where a single wave function enters eq.(7), the
resulting
equations can be more conveniently rewitten by introducing the modulus
and the phase of the wave function
$$
\Psi = \sqrt{\rho} \ e^{iS} \ \ \ .
\eqno(9)
$$
Here $\rho$ is the (diagonal) density of the system. The resulting equations
then take the form of the equations of hydrodynamics:
$$
{\partial \over \partial t } \rho + \nabla({\bf v}\rho) = 0
\eqno(10)
$$
$$
{\partial \over \partial t } S + {\delta \over \delta \rho} E = 0
\eqno(11)
$$
where ${\bf v} = {1\over m}\nabla S$ is the velocity field and $S$ has the
meaning of a velocity potential.
\vskip 28 truept

\centerline{\bf 2. TIME DEPENDENT DFT AND RPA}
\vskip 12 truept

In the bulk the equations of TDDFT can be easily employed to calculate the
linear response of the system. Let us consider, for simplicity, the
density-density response function and let us assume for the moment
(see however the discussion in the second part of the work) that
the functional depends, apart from the kinetic energy term, only on the
diagonal
density $\rho$:
$$
{\cal H} = -\Psi_i^* {1\over 2m} \nabla^2 \Psi_i + V(\rho)
\eqno(12)
$$
The linear response function can then be easily evaluated
in this case and takes the familiar form
$$
\chi(q,\omega) = {\chi_0(q,\omega) \over 1 - v(q) \chi_0(q,\omega)} \ \ \ ,
\eqno(13)
$$
of the random phase approximation (RPA). In eq.(13) $\chi_0(q,\omega)$ is
the independent particle response function given by the Lindhard function for
fermions and by the expression
$$
\chi_0(q,\omega) = {q^2 \over 4m} \ {1 \over \omega^2 - (q^2/2m)^2}
\eqno(14)
$$
for bosons. The quantity $v(q)$ is the Fourier transform of $\delta^2 V(\rho)/
\delta\rho({\bf r}_1)\delta\rho({\bf r}_2)$ and is entirely fixed by the static
response of the system
$$
\chi(q) \equiv \chi(q, \omega\!=\!0) =
{\chi_0(q) \over 1 - v(q) \chi_0(q)}\ \ \ .
\eqno(15)
$$
For electrons the quantity $v(q)$ is usually written in the form
$v(q) = 4 \pi e^2 q^{-2} (1-G(q))$, where $G(q)$ is the so called local field
correction. Notice that despite the fact that eq.(13) holds for both bosons and
fermions, the resulting
structure of $\chi(q,\omega)$ and in particular its poles
are deeply different in the two cases as a consequence of the different form of
$\chi_0$. Equation (13) can be  easily generalized to include multi-component
systems (for example $^3$He-$^4$He mixtures).

The above discussion reveals that the structure of TDDFT (in its
linearized form) is identical to the RPA. These theories are mean field
theories
and are able to provide an adequate picture of elementary excitations in the
low $q$, low $\omega$ regime. Actually in this regime they coincide
with the Landau theory of Fermi liquids in the case of fermions and with the
equations of classical hydrodynamics in the case of bosons. These are the
proper
theories to describe macroscopic phenomena in strongly interacting quantum
liquids. at zero temperature.

In conclusion we can say that TDDFT is well suited to investigate the
elementary
excitations of quantum many body systems at least in the macroscopic
regime. However, due to its mean field nature,
this theory cannot account for multi-pair excitations and consequently
{\bf cannot} provide a complete description of $S(q,\omega)$. The situation
is schematically drawn in fig.3,
where we have distinguished between the low $\omega$ region dominated by
elementary excitations (collective modes and, in Fermi systems, single
particle transitions) and a higher $\omega$ region dominated by
multi-pair effects.
\topinsert
\vskip 10 truecm
\noindent
{\bf Figure 3.} Schematic representation of $S(q,\omega)$. For bosons
one should ignore
the part corresponding to single particle excitations.
\vskip 28 truept
\endinsert

In  deriving  result (13) for the dynamic response function we have made the
assumption that the interaction terms in the energy functional depend only on
the diagonal density (see eq.(12). This is clearly an approximation which
permits to express the
{\it dynamic response} in terms of the {\it static response} (see eqs.(13-15)).
 In the language of the Landau theory of Fermi liquids this
corresponds to assuming the Landau parameter $F_{\ell}$ is equal to zero when
 $\ell \ge 1$.
Is this approximation correct enough? The answer is in general negative
in strongly interacting liquids and we well know that in liquid $^3$He the
dynamic response has a form which differs from eq.(13) due to the
presence of the Landau parameter $F_1$. In the language of DFT this means that
one cannot ignore the occurrence of current interaction terms in the energy
functional. In liquid $^3$He this effect has the important consequence of
fixing the difference between the first sound ($c_1$) and the zero sound
($c_0$)  velocity according
to the formula (valid within minor approximations):
$$
c^2_{1} = c^2_{0} + {4\over 15}{p_F^2 \over m^2(1+F_1^s/3)}
\eqno(16)
$$
The experimental confirmation of this difference has been shown to be
consistent with the
measured value of the effective mass
$$
{m^*\over m} = 1 + {1\over 3} F_1^s
\eqno(17)
$$
thereby providing a direct check of the correctness of the Landau theory.

\vskip 28 truept

\centerline{\bf 3. SUM RULES AND MULTI-PAIR EXCITATIONS}
\vskip 12 truept

Figure 3  explicitly reveals that if  one evaluates suitable
moments of the dynamic structure factor
$$
m_k = \int_0^{\infty} \! d\omega \ \omega^k S(q,\omega) \ \ \ ,
\eqno(18)
$$
the results obtained using TDDFT will differ
from the ones derivable from an exact
calculation or, in principle, from experiments due to the role played
by multi-pair excitations which give a non vanishing contribution
to the integrals   (18). This question is relevant
because  in some cases these moments can be evaluated with the help of
sum rule techniques. Many of these sum rules play an important role in
the physics of the many body problem (see for example the f-sum rule)
and it is consequently important to understand the relation between sum rules
and the TDDFT-RPA scheme.

Due to the occurrence of multi-pair excitations in general mean field theories
do not fulfill the sum rules even at small $q$. This statement is important
because, for example,
one usually expects the Landau theory of Fermi liquids to be exact at
small $q$. This
is true only if one limits the calculation of the dynamic structure factor
to the low
$\omega$ regime too. In general the integrals (18) take a significant
contribution from the high $\omega$-region, a region where the mean field
theory  cannot provide a proper description.

The inverse energy weighted moment $m_{-1}$ yielding the static response (see
eq.(5)) is an exception in this sense. In fact in the low $q$
limit the multi-pair
contribution to this moment is vanishingly small and $m_{-1}$ is consequently
exhausted by the collective excitation and (in Fermi systems) by single
particle excitations. This property of $m_{-1}$ is especially remarkable in
connection with the fact that the static response is exactly reproduced by DFT.
It ensures a beautiful self-consistent behavior of the theory.

A different situation occurs with the other moments of the dynamic
structure factor.
An important question to discuss in this context is whether the $f$-sum rule
$$
\int_0^{\infty} d\omega\omega S(q,\omega) = {q^2 \over 2m}
\eqno(19)
$$
is exhausted by the elementary excitations of the system.
Result (19) follows from the use of the completeness relation and
holds for velocity independent potentials. Differently from
the inverse energy weighted moment the answer to the above question is
in general  negative even at small $q$.
A well known example confirming this statement is given by spin excitations
in liquid $^3$He [6] where the Landau theory, which properly accounts for
the low  $q$, low $\omega$ behavior of the spin dynamic structure factor,
gives the result
$$
\int_0^{\infty} \! d\omega\omega \ S^L(q,\omega) =
{q^2 \over 2m} {1+F_1^a/3 \over 1+F_1^s/3}
\eqno(20)
$$
a quantity significantly smaller than the exact result (19) holding for
spin independent interatomic potentials. A similar situation occurs in
antiferromagnets
as well as in bosonic systems in a lattice or in the presence of disorder.
The general rule is that {\sl the energy weighted sum rule is not exhausted,
at small $q$, by the
elementary excitations of the system in those systems where the current
is not conserved. This happens in general for spin excitations and also for
density excitations if the system is not translationally invariant}.
\vskip 28 truept
\centerline{\bf 4. SUPERFLUID $^4$He}
\vskip 12 truept

Since the current is conserved in $^4$He the f-sum rule is
entirely exhausted by
the phonon mode at small $q$ in agreement with the general statements
of classical hydrodynamics. At higher $q$ the relative
contribution of the elementary excitations to the
integral (19) becomes smaller and smaller and is about  one third in the
rotonic
region as a consequence of the important role played by multi-pair
excitations at high wave vectors.
Viceversa the inverse energy weighted sum rule even at relatively
large $q$ is mainly dominated by the elementary mode due to the occurrence
of the $\omega^{-1}$ factor in the integral (5) which quenches
the contributions arising from multi-pair excitations located at higher
energies.

Let us discuss the implications of the above discussion in the case of $^4$He.
In the absence of current interaction terms in the energy functional the
dynamic
response function takes the form (see eqs.(13-15)):
$$
\chi(q,\omega) = {q^2 \over 2m} {1\over \omega^2 - \omega^2_0(q)}
\eqno(21)
$$
and exhibits a single pole whose dispersion law is given by
$$
\omega_0^2(q) = {q^2 \over 2m} \left(  {q^2 \over 2m}+2v(q) \right) \ \ \ .
\eqno(22)
$$
\topinsert
\vskip 10 truecm
\noindent
{\bf Figure 4.} Dispersion of elementary excitations in superfluid $^4$He.
Full line: Feynman approximation. Dashed line TDDFT without current terms
(eq.()). Points: experimental data (from [7]).
\vskip 28 truept
\endinsert

In fig.4 we report the dispersion law (22). The function $v(q)$
is fixed, through eq.(15), by the static response function which is known
experimentally in superfluid $^4$He. The resulting curve for $\omega(q)$
overestimates the
experimental curve.  Actually result (22) can be also
rewritten in the form
$$
\omega^2_0(q) = {m_{1}(q) \over m_{-1}(q)} = -{q^2\over m \chi(q)}
\eqno(23)
$$
which corresponds to a rigorous upper bound to the dispersion of the collective
branch (we ignore here possible decay mechanisms of elementary excitations).

It is interesting to compare result (23) with the prediction of the most
famous  Feynman
approximation
$$
\omega(q) = {m_1(q)\over m_0(q)} = {q^2 \over 2mS(q)}
\eqno(24)
$$
which expresses the energy of the elementary excitation in terms of the
static structure factor  $S(q)$.
The Feynman result (24) also provides a rigorous upper
bound to the exact dispersion
law. It is important to remark that both the bounds (23) and (24)
coincide with the phonon dispersion $\omega = cq$ at small $q$. One has
the general inequality
$$
\sqrt{{m_1\over m_{-1}}} \le {m_1\over m_0}\ \ \ .
\eqno(25)
$$
Inequality (25) reveals that, even in the absence of current terms
in the energy functional, time dependent
DFT results to be a better approximation with respect to Feynman theory,
providing  an upper bound closer to the exact dispersion.
{}From a microscopic point of view the quantity $\chi(q)$ is  much more
difficult to calculate with respect to $S(q)$ though first caculations of
$\chi(q)$,
based on diffusion Monte Carlo techniques, are now becoming available [8].

We are now ready to discuss the inclusion of current dependent terms in DFT.
The explicit form of these new terms is discussed in ref.[2] and will not
reported here. The new term is included phenomenologically with the criterium
of reproducing the experimental dispersion law of elementary excitations.
The following remarks are in order here:
\smallskip
\item{$\bullet$} The new term results in  a modification of the equation of
continuity (10) which now contains an extra contribution having the form of a
backflow effect. Viceversa the Euler equation keeps its form (11).
\smallskip
\item{$\bullet$} The new term does not affect the Galilean invariance and
in particular
the low $q$ behavior of the response function is not modified. Important
changes
are instead introduced at higher $q$ especially in the roton region.

\smallskip

 The formalism of TDDFT is now ready to be applied
to study the dynamics of inhomogeneous  helium  systems.
One can explore, for example, the interesting
region where surface excitations can  couple with
rotons (ripplon-roton hybridization)[9] and where elementary
excitations can give rise
to particle emission (quantum evaporation)[10].
These investigations are the object of
present research and will hopefully provide a test of the quality of the TDDFT
formalism and in particular of the new density functional.

Finally one should also be also aware  of the limits of the theory.
In particular
the present formalism cannot account for damping effects associated
with the decay of an elementary excitation into two or more excitations.
Furthermore the formalism cannot be pushed upto very high momenta (larger
than about
$2.5 A^{-1}$). In fact at these high values of $q$ the inverse energy weighted
moment (5) is no longer
exhausted by the elementary excitations: in this regime the static response,
major ingredient of DFT, has lost its connection with the physics of elementary
excitations.

In conclusion the main message emerging from the present discussion is that the
theoretical basis of TDDFT (with the inclusion of current dependent terms)
is now reasonably well established and can be  applied in a reliable
way to investigate
important dynamic phenomena of inhomogeneous systems in the microscopic regime.

\vskip 28 truept

\centerline{\bf REFERENCES}
\vskip 12 truept

\item{[1]} Note that $\hbar=1$ in this work.

\item{[2]} F. Dalfovo, A. Lastri, L. Pricaupenko, S. Stringari, and J.
Treiner, in preparation

\item{[3]} J. Boronat, J. Casulleras, and J. Navarro, preprint
1994.

\item{[4]} B.M. Abraham, Y. Eckstein, J.B. Ketterson, M. Kuchnir,
and P.R. Roach, Phys. Rev. A {\bf 1}, 250 (1970).

\item{[5]} R.A. Cowley and A.D.B. Woods, Can. J. Phys. {\bf 49},
177 (1971); A.D.B Woods and R.A. Cowley, Rep. Prog. Phys. {\bf 36}, 1135
(1973)

\item{[6]} Landau theory

\item{[7]} R.J. Donnelly, J.A. Donnelly and R.N. Hills, J. Low
Temp. Phys. {\bf 44}, 471 (1981).

\item{[8]} S. Moroni, D. M. Ceperley, and G. Senatore, Phys. Rev.
Lett. {\bf 69}, 1837 (1992); S. Moroni, private communication.

\item{[9]} A. Lastri, F. Dalfovo, L. Pitaevskii, and S. Stringari, J.
Low Temp. Phys., in press

\item{[10]} F. Dalfovo, A. Lastri, L. Pitaevskii, and S. Stringari,
in preparation

\end